# Critical Data Studies in the Anthropocene

Ana Valdivia

*This chapter introduces the concept of the Anthropocene into critical data studies, a field that has, for the past decade, explored the entanglements between datafication and politics. With the scaling up of contemporary datafication alongside generative computing, it is essential to expand these debates, both theoretically and methodologically, to consider the logics of extraction and exploitation inherent in the politics of artificial intelligence. Critical data scholars are called to interrogate how dominant narratives around the materiality of contemporary datafication either obscure or reveal its environmental impacts, along other key questions such as who benefits from these logics. To illustrate this, this chapter brings two case studies from Spain and Chile and explores how the infrastructure of contemporary datafication intersects with existing power structures, influencing which social and environmental consequences are recognized, addressed, and neglected. Finally, it invites critical data scholars to keep exploring the politics of data, infrastructure, and the Anthropocene by blending discipline boundaries between science and technology studies, media, geography, and political ecology.*

"Los bosques, ríos y cerros no son cosas. Tampoco lo son las plantas, los animales y la diversidad de seres de la naturaleza. [...] No son seres para vender y explotar para lucrar."

*Forests, rivers and hills are not things. Neither are plants, animals and the diversity of nature's beings. [...] They are not beings to be sold and exploited for profit*

Elisa Loncon (*Azmapú*) 2023.

The concept of the *Anthropocene*—from the ancient Greek *anthropos* "human" and *-cene* "recent"—was coined to describe a new geological epoch in which humans are altering nature.[1] Given scientific observations brought by researchers that noted composition changes in the atmosphere, water bodies, and soil since the industrialization era, they proposed to establish this concept to reflect how humans have become a new geophysical force. Nuclear bombs, steam engines and coal industries have had an impact on the planet that was evident through scientific observations. Given its relevance, the Anthropocene rapidly became adopted by scholars from other disciplines, such as political economy, geography, and science and technology studies (STS), demonstrating that these changes observed on the world-ecology system due to industrial and economic activities stimulated different analytical perspectives. As a result, several conceptual alternatives have been proposed to the Anthropocene. Critical political economist scholars Andrea Malm and Jason Moore proposed the *Capitalocene* as "[it] signifies capitalism as a way of organising nature".[2] Their point is that not every human contributes equally to the Anthropocene, but capitalism as a world-ecology of power, capital, and nature does. From an STS perspective, Haraway proposed the *Chthulucene*, noting the interlinked and tentacular connections between humans and non-humans in the web of life.[34] Media ecologist Jussi Parikka proposed the *Anthrobscene* arguing that "the addition of the obscene is self-explanatory when one starts to consider the unsustainable, politically dubious, and ethically suspicious practices that maintain technological culture and its corporate networks."[5] Despite these interdisciplinary approaches, the critical data literature has offered a rather limited engagement with these debates on the Anthropocene.

[1] Paul Crutzen and Eugene Stoermer, "The 'Anthropocene,'" *Global Change Newsletter* 41 (2000): 17–18.

Given this gap within critical data studies, I propose to discuss the materiality of contemporary datafication and generative computing technologies, which have significantly amplified their contribution to the Anthropocene and added dramatic changes in the world-ecology system. Traditionally, critical data studies have analyzed power structures within datafication mechanisms. However, this literature is disconnected from media infrastructure studies, that have largely investigated the "material forms as well as discursive construction" of the movement of electronic media throughout data centers, subsea cables and telecommunication towers.[6] The collection of data through digital devices together with the development of advanced semiconductors have enabled the design of the contemporary *attention/transformer based-use of internet-scale training datasets*.[7] These algorithms are label as "generative" models because *generate* content such as text, images, and videos. However, the increasing computational demands of these models, combined with their popularity in capital markets and digital services, have significantly expanded their user base. Consequently, the resources that are necessary to power this technology have increased, and the materiality supporting generative computing has become a focal point for critical academic inquiry in the current climate crisis. Generative computing contributes to climate warming by emitting more greenhouse gases, extracting more water, and occupying more land, which degenerates natural landscapes, human and nonhuman beings. While critical data scholars have hesitated to engage with the Anthropocene, the contribution of this chapter to the *Handbook of Critical Data Studies* is an invitation to rethink this field by continuing to blur disciplinary boundaries, both theoretically and methodologically. It calls for applying our critical lens to the Anthropocene and its relationship within infrastructural assemblages and contemporary datafication.

---

[2] Jason W. Moore, ed., *Anthropocene or Capitalocene? Nature, History, and the Crisis of Capitalism* (Oakland, CA: PM Press, 2016), 6.

[3] Donna J. Haraway, *Staying with the Trouble: Making Kin in the Chthulucene* (Durham, NC: Duke University Press, 2016).

[4] Sy Taffel, *Digital Media Ecologies: Entanglements of Content, Code and Hardware* (London: Bloomsbury Academic, April 22, 2021), 2.

[5] Jussi Parikka, *The Anthrobscene* (Minneapolis: University of Minnesota Press, 2014), 10.

[6] Lisa Parks and Nicole Starosielski, eds., *Signal Traffic: Critical Studies of Media Infrastructures* (Chicago, IL: University of Illinois Press, 2015), 7.

## Datafication, Infrastructure, and Power in the Anthropocene

Critical data studies have investigated how power dynamics are recrafted through datafication mechanisms by exploring the "unique cultural, ethical, and critical challenges posed by Big Data".[8] The use and accumulation of data have been instrumentalized to impose political ideologies and truth regimes through scientific objectivity. Yet critical data scholars have shown that data is inherently a form of politics, thus is not a neutral tool. Dalton and Thatcher, who coined the concept of critical data studies, argued that data, as a technology, is a political, historical and geographical phenomenon that embeds epistemological power and invites scholars to "critically ask who it speaks for".[9] Since then, critical data scholars have interrogated how data and algorithms are instrumentalized to

[7] I attribute this expression to Zac Zimmer (University of California).

[8] Andriana Iliadis and Flavia Russo, "Critical Data Studies: An Introduction," *Big Data & Society* 3, no. 2 (2016): 1, https://doi.org/10.1177/2053951716674238.

[9] Craig Dalton and Jim Thatcher, "What Does a Critical Data Studies Look Like, and Why Do We Care? Seven Points for a Critical Approach to 'Big Data,'" *Society & Space* (May 12, 2014), n.p., https://www.societyandspace.org/articles/what-does-a-critical-data-studies-look-like-and-why-do-we-care

reproduce exploitative systems and the cultural and political values of domination.[10,11,12] In a similar vein, David Beer, builds on Foucault's work to eloquently expose how algorithmic systems, which are in turn trained through datafication mechanisms, are recrafting forms of power.[13] Beer argues that algorithmic systems encapsulate regimes of truth provided by historical social ordering processes.[14] While datafication is not a recent phenomenon, in fact governments and institutions have been collecting data on racialized subjects for decades[15], the introduction of digital devices has accelerated its accumulation and facilitated surveillance mechanisms. Our societies are witnessing a growing volume of data year by year, stored in servers within data centers and cloud networks. Combined with advances in the semiconductor industry, the field of artificial intelligence has developed attention and transformer-based architectures that generate content by scraping data from the Internet, often reproducing hegemonic values.[16,17] Yet, while geographers and anthropologists have extensively analyzed the power dynamics embedded in other types of infrastructure, the current critical data literature has yet to provide a theoretical framework addressing the materiality of generative technologies.18,19

Human geographer Graham Pickren analyzed the circulation of power through "infrastructural assemblages" of big data.[20] Pickren pointed out that, like data and algorithms, their infrastructure is not neutral, and it has been metaphorically obfuscated through "the cloud"— resonating with Parikka's reflection on the *Anthrobscene*.[21] Through these obfuscation mechanisms, the infrastructure of technology and its commodities become fetishized.[22] Questions surrounding labor and environmental rights of these commodities, such as where these servers have been manufactured or which territories might be sacrificed to train the next transformer algorithm, remains unaddressed. Infrastructural power is exercised by rendering invisible the logics of extraction and exploitation involved in the making of artificial intelligence. However, they do remain visible to those suffering the consequences of computational infrastructure.

[10] Simone Browne, "*Dark matters: On the surveillance of blackness,"* Durham: Duke University Press, (2015).

[11] Jonas C. L. Valente and Rafael Grohmann, "Critical Data Studies with Latin America: Theorizing beyond Data Colonialism," *Big Data & Society* 11, no. 1 (2024), 10.

[12] Ruha Benjamin, "*Race after Technology: Abolitionist Tools for the New Jim Code,"* Oxford: Polity (2019).

[13] David Beer, "The Social Power of Algorithms," Information, Communication & Society 20, no. 1 (2017), 8.

[14] Beer, 1.

[15] Ana Valdivia and Martina Tazzioli, "Datafication Genealogies beyond Algorithmic Fairness: Making Up Racialised Subjects," in *Proceedings of the 2023 ACM Conference on Fairness, Accountability, and Transparency* (Chicago, IL: Association for Computing Machinery, June 12–15, 2023), 847.

[16] Karen Hao, *Empire of AI: Dreams and Nightmares in Sam Altman's OpenAI* (New York: Penguin Press, May 20, 2025), 15.

[17] Louise Amoore, Alexander Campolo, Benjamin Jacobsen, and Ludovico Rella, *"A World Model: On the Political Logics of Generative AI," Political Geography* 113 (2024).

[18] Valdivia, 2025. Pieper, 2025.

[19] Signe Sophus Lai and Sofie Flensburg, "Scaling up: directions for critical studies of data infrastructure. In T. Venturini, A. Acker, J. Plantin, T. Walford (Eds.) *The Sage Handbook of Data and Society, (2025).*

[20] Graham Pickren, "The Global Assemblage of Digital Flow: Critical Data Studies and the Infrastructures of Computing," *Progress in Human Geography* 42, no. 2 (2018), 226.

As transformer algorithms increasingly capture the attention of economic markets, [23] their growing scale and complexity have brought attention to the substantial resources needed for their training. The scale of these algorithms has revealed Pickren's infrastructural assemblages and its epistemological power. Power is not only embedded in the contentious data extraction practices but also in the material and environmental resources extracted.[24] Power operates through political decisions that determine economic growth should rely on greater computational capacity. Power operates through the politics of urban planning, shaping decisions about where to site data centers and facilitating the deployment of infrastructural processes. Power further renders opaque the supply chains and actors that orchestrate these infrastructures, while also obfuscating the environmental consequences. The more data is generated and consumed, the more nature is degenerated through power mechanisms that renders invisible how the materiality of artificial intelligence is contributing to the Anthropocene.

## The Materiality of Artificial Intelligence

[21] Pickren, 229.

[22] Ana Valdivia. "Follow The Thing AI" in *AI Infrastructures and Sustainability* eds. Anne Mollen, Sigrid Kannengießer, Fieke Jansen, Julia Velkova (Münster: Palgrave, 2025), 15.

[23] Emily M. Bender and Alex Hanna, *The AI Con: How to Fight Big Tech's Hype and Create the Future We Want* (New York: Harper, 2025), 22.

[24] Ana Valdivia, "The Supply Chain Capitalism of AI: A Call to (Re)Think Algorithmic Harms and Resistance through Environmental Lens," *Information, Communication & Society*, October 2024, 3. Maximilian Pieper, "Is data material? Toward an environmental sociology of AI," *AI & SOCIETY*, 1-12, July 2025.

According to anthropologist Elana Resnick, materiality is "the study of objects and their relationships to and in social life."[25] The study of the materiality of artificial intelligence entails the investigation of the objects related to this technology together with its social, political, and economic values.[26] Data centers, wires, subsea cables and chips are infrastructural elements that made the materiality of AI that are manufactured, built and deployed across geographies, transforming electricity into data. In this vein, STS and geographer scholar Ludovico Rella argues that "materiality […] is often used to 'ground' digitality, or to show how in both material and immaterial, analogue and digital environment, big data industry retains the same extractive logic: extraction of minerals, extraction of data".[27] Drawing on Rella's argument, I argue that the materiality of artificial intelligence is a key theoretical framework for critical data scholars to unveil the connection between generative computing and the Anthropocene.[28]

However, a closer examination of the materiality of artificial intelligence reveals both continuities and discontinuities. Media ecologist Sy Taffel wrote in 2023 that "the production of data is always a material encounter", adding that "digital technologies are materially complex artefacts, typically requiring between 60 and 70 of the 84 nonradioactive elements found on earth".[29] Attention and transformer algorithms, as a type of digital technology, are not an exception within this material encountered. The key material component needed to train these algorithms are graphic processing units.[30] These chips have enabled the training of increasingly deep learning infrastructures, but the volume of minerals required to manufacture these units has grown over time. This fact could entail

---

[25] Elana Resnick. "Materiality", *Oxford Bibliographies Online*, last modified February 21, 2022, accessed June 6, 2025, https://www.oxfordbibliographies.com/display/document/obo-9780199766567/obo-9780199766567-0277.xml.

[26] Luke Munn, "Red Territory: Forging Infrastructural Power," *Territory, Politics, Governance* 11, no. 1 (2023): 81. Maximilian Pieper. "Is data material? Toward an environmental sociology of AI," *AI & Society*, (2025): 1-12. Federico Cugurullo, et al. The nature of AI: Metabolism, energy, water, labour and justice in the urban political ecology of artificial intelligence. *Urban Political Ecology 1(1-2),* (2025), 33-53.

[27] Rella, 6.

[28] Jathan Sadowski, "When Data Is Capital: Datafication, Accumulation, and Extraction," *Big Data & Society* 6, no. 1 (January 7, 2019): 1.

[29] Sy Taffel, "Data and Oil: Metaphor, Materiality and Metabolic Rifts," *New Media & Society* 25, no. 5 (2023), 984.

[30] Ludovico Rella, "Close to the Metal: Towards a Material Political Economy of the Epistemology of Computation," *Social Studies of Science* 54, no. 1 (February 2024): 3–29

that more territories are going to be sacrificed to extract the minerals to manufacture the materiality of artificial intelligence.

But it is not only about the diversity of materials needed, it is also about 'scale'. Compared to machine learning algorithms trained just a decade ago, the materiality of today's artificial intelligence needs larger data centers and more resources, including fossil fuels, water, and land.[31] As an example, Meta used 22 million liters of water in 97 days to train Llama-3. This is the same amount of water a person in London would use in more than 400 years.[32] As another example, in 2023, Google increased its carbon emission by 13% due to artificial intelligence energy demand.[33] Therefore, the economic growth of artificial intelligence industry is translated into a large environmental footprint given the number of natural resources needed to train its models.

Given this context, critical data scholars could also interrogate in what ways do dominant narratives around the materiality of artificial intelligence obscure or reveal the environmental and political effects and who benefits from this framing? How does the materiality of AI intersect with existing power structures to influence which environmental impacts posed by intense datafication and large computational models are recognized, addressed, or ignored? And more importantly, how could critical data scholars navigate the logics of extraction and exploitation featured within the politics of the Anthropocene "without embracing arrogance, universalism, and naïve materialism"?[34]

[31] Pengfei Li, Jianyi Yang, Mohammad A. Islam, and Shaolei Ren, "Making AI Less 'Thirsty': Uncovering and Addressing the Secret Water Footprint of AI Models," *arXiv* preprint, April 6, 2023, https://arxiv.org/abs/2304.03271.

[32] Ana Valdivia, "Data Ecofeminism," in *Proceedings of the 2025 ACM Conference on Fairness, Accountability, and Transparency (FAccT '25)*, Athens, Greece, June 23–26, 2025 (New York: Association for Computing Machinery, 2025), 2.

[33] Dan Milmo, "Google's Emissions Climb Nearly 50% in Five Years Due to AI Energy Demand," *The Guardian*, July 2, 2024, https://www.theguardian.com/technology/article/2024/jul/02/google-ai-emissions.

[34] Penny Harvey, Christian Krohn-Hansen, and Knut G. Nustad, eds., *Anthropos and the Material* (Durham, NC: Duke University Press, 2019), 3.

## The Environmental Impacts of Contemporary Datafication: Metrics, Obfuscation, and Miscalculations

How is the impact of generative computing in the Anthropocene quantified nowadays? While big tech companies, such as Meta, Google, and Microsoft are not legally obliged to inform about their environmental impacts, they annually publish Corporate Sustainability Reports (CSRs hereinafter) in which they promote their environmentally responsible strategies and practices. A close reading of these reports reveals self-reported metrics about, for instance, carbon emissions of Google's data centers and water usage of Microsoft's cloud. Yet, the mechanisms for accounting for these metrics are not standardized, meaning there is no established methodology to rigorously report them on annual basis. Consequently, these reports have been largely criticized for being financially performative through greenwashing narratives, with critical scholars advocating against the self-regulation of the big tech industry in the context of sustainability.[35] Consider the number of private actors within the supply chains of artificial intelligence who fail to publicly and independently report their environmental metrics and obfuscate these metrics through sustainability accountability mechanisms.

According to an investigation published by The Guardian, the big tech industry could probably contribute to 662% higher carbon emissions that they reported.[36] Within CSRs, carbon emissions are quantified in three categories: Scope 1, Scope 2, and Scope 3. Scope 1 refers to direct emissions from company-owned or controlled sources, such as emissions from their vehicles and on-site operations. Scope 2 accounts for emissions from the generation of purchased electricity consumed by the company, typically calculated based on energy provider data. Scope 3 includes all other indirect emissions across the company's supply chain, such as those from suppliers and product use.

---

[35] Photini Vrikki, *Measuring Up? The Illusion of Sustainability and the Limits of Big Tech Self-Regulation,"* *Sustainability* 16 (2024): 11.

[36] Isabel O'Brien, *"Data center emissions probably 662% higher than big tech claims. Can it keep up the ruse?" The Guardian*, September 15, 2024, https://www.theguardian.com/technology/2024/sep/15/data-center-gas-emissions-tech.

Large companies such as Google and Microsoft are required to report Scope 1 and Scope 2 emissions. Within these environmental metrics and categories there are further concepts to take into account. Companies report Scope 2 emissions by dividing them into location-based and market-based categories. However, market-based emissions may be misleading, as they might not accurately reflect the true carbon emissions associated with a company's energy consumption. The reason is that market-based emissions include Renewable Energy Certificates (RECs), an engineering accounting mechanism that discounts carbon emissions if companies buy renewable energy. Yet, according to academics and practitioners, RECs have been considered *greenwashing* mechanisms as these certificates only involve paying an extra fee while not having any change in overall carbon emissions.[37] Therefore, Scope 2 market-based emissions are miscalculating emissions that could have serious consequences in the contribution of big tech industry to climate warming. As a result, market-based emissions metrics are often instrumentalized by the tech industry to account for their environmental impacts. However, academics, journalists, stakeholders, and policymakers must remain critical of these obfuscation strategies to avoid falling into the trap of misleading accounting practices.

Despite these engineering and accounting mechanisms, the big tech industry continues to increase its emissions. In 2025, Google announced that emissions have soared by 51% since 2019, underscoring the environmental costs associated with the generative computing race.[38] This surge in emissions is largely driven by the increasing computational demands of artificial intelligence, a key driver of contemporary datafication. Many critical scholars have argued that, while companies like Google frame their green initiatives as part of a larger sustainability effort, these efforts often reflect greenwashing rhetoric rather than genuine progress. Such narratives frequently fail to address the true ecological footprint of artificial intelligence and their substantial contribution to the Anthropocene, where nature is increasingly destroyed to maintain economic growth and foster computational innovation.[39,40] An important consideration when accounting for carbon emissions is that, in most cases, such emissions are measured only during the training phase of the technology, specifically, within the data center. However, the materiality of AI and its contribution to the Anthropocene in the form of carbon emissions extends beyond data center operations to include other activities such as mineral extraction, transportation, and chip manufacturing. As a result, emissions generated throughout the broader AI supply chain are often excluded from CSRs. While technically these emissions should be accounted for under Scope 3, assigning accounting responsibilities across supply chains is a complex task. First, companies often do not hold their suppliers or service providers accountable for emissions. Second, the orchestration of multiple actors within a supply chain makes it particularly challenging to comprehensively account for all associated carbon emissions.

---

[37] Brander et al. *Open Letter Rejecting the Use of Contractual Emission Factors in Reporting GHG Protocol Scope 2 Emissions," Scope 2 Open Letter*, February 12, 2015, https://scope2openletter.wordpress.com/2015/02/12/open-letter-rejecting-the-use-of-contractual-emission-factors-in-reporting-ghg-protocol-scope-2-emissions/#ftn2.

[38] Helen Horton, *"Google's Emissions up 51% as AI Electricity Demand Derails Efforts to Go Green" The Guardian*, June 27, 2025, https://www.theguardian.com/technology/2025/jun/27/google-emissions-ai-electricity-demand-derail-efforts-green

As I argue in this chapter, critical scholars must also recognize that environmental data is inherently subjective and can be manipulated through technical or engineering mechanisms or simply obscured due to the opacity of global supply chains, thereby concealing the full extent of environmental impacts. I argue that critical data studies should remain positioned as a field that challenges the presumed neutrality of data by revealing how processes of counting and analysis are not neutral, neither in their social or environmental contexts.[41]

## Beyond Environmental Metrics: Data Centers and More-than-Humans

While the previous section illustrates how the tech industry obfuscates environmental metrics through engineering and financial mechanisms, this section illustrates other mechanisms that I have encountered during my fieldwork. The two case studies presented reveal how the big tech industry addressed environmental critique by local communities before data centers were built. The first case presents Meta's hyperscale project in Talavera de la Reina, Spain. While environmental and new against-data-centers grassroots organizations have raised concerns against this project, Zuckerberg's company is going to deploy its project in the incoming years. The impact of this data center will extend beyond human lives, affecting the eagles and vultures that use that Spanish region for nesting and feeding. The second case focuses on Amazon's data center project in Huechuraba, Chile. It reveals that the local community was less concerned with the construction of the data center itself and more with the surrounding infrastructure, that is high-voltage towers or pylons, that would be built across a nearby hill to meet the center's electricity demand. Although these two cases are situated in different geographical locations and operate within distinct economic and political contexts, their empirical analysis reveals how both companies obscure their environmental metrics beyond what is disclosed in their CSRs to address local communities' criticism.

---

[39] Mél Hogan and Gwendolyn Blue, *"Big Cloud Solastalgia,"* in *Digital Technologies for Sustainable Futures: Promises and Pitfalls*, ed. Chiara Certomà, Fabio Iapaolo, and Federico Martellozzo (London: Routledge Taylor & Francis Group, 2025), 32–45.

[40] Patrick Brodie, *"Climate Extraction and Supply Chains of Data," Media, Culture & Society* 42, no. 7-8 (2020): 1099.

[41] David Beer, *"Envisioning the Power of Data Analytics," Information, Communication & Society* 21, no. 3 (2023): 465–79.

## Data Infrastructure, Eagles and, Vultures in rural Spain

Talavera de la Reina is a town located in Castilla-La Mancha, a Spanish region widely known for its rural legacy. However, this area has experienced severe depopulation due to sustained migratory flows from rural to urban areas, a phenomenon often referred to as "emptied Spain" (*España vaciada*).[42] Talavera de la Reina has recently gained attention as the site of Meta's first hyperscale data center in Spain. The project has been classified as of "singular interest" by the regional government of Castilla-La Mancha, alongside a casino, an airport, and a golf course. This designation facilitates urbanization through land reclassification and allows developers to bypass environmental and urban planning regulations.[43]

The data center will occupy 191 hectares and includes plans to build 130,000 square meters of data farms. Meta and its partner Zarza Networks present the project as environmentally sustainable, promising actions such as "reversing biodiversity loss" and "restoring more water volume than is consumed".[44] According to their environmental report, the data center will have an installed electrical capacity of 248 MW, equivalent to the annual electricity consumption of approximately 71,000 Spanish households. Its estimated drinking water usage is 327 million liters per year, comparable to the annual water consumption of 7,000 households in the area.

[42] Carlos Taibo, *Iberia vaciada: Despoblación, decrecimiento, colapso* (Madrid: Los Libros de la Catarata, 2021), 4.
[43] Luis Alfonso Escudero-Gómez, "*Construir cualquier cosa en cualquier lugar: los Proyectos de Singular Interés en la región de Castilla-La Mancha (España)*" *EURE (Santiago)* 49, no. 147 (2023): 1.

The report also acknowledges that the project is located within the designated area of the Recovery Plan for the Iberian Imperial Eagle and the Black Vulture. However, it claims that the land used by the data center accounts for only 0.004% and 0.009% of the relevant zones for these species, respectively. To assess the impact on protected species, the developers conducted bird counts during six fieldwork visits between December 2021 and June 2022.

However, Meta's environmental report has faced criticism from environmental organizations, academics, and journalists. One of the primary concerns raised was the projected water consumption—an especially sensitive issue given Spain's ongoing and severe droughts. A formal critique by SEO/BirdLife, Spain's leading ornithological organization, focused on the excessive water usage associated with the facility. Although this concern was initially dismissed, Meta later revised its estimates, reducing the data center's projected drinking water consumption from 327 million to 40 million liters annually. This revision also lowered the estimated peak water demand from 37 to 10 liters per second. Nonetheless, even at these reduced levels, Meta's data center is expected to consume approximately 8% of the total water usage of Talavera de la Reina.[45]

SEO/BirdLife also contested the adequacy of the bird monitoring methodology used in the environmental assessment. The organization recommended extending the study period to at least one year to allow for a more comprehensive evaluation of the data center's impact on local wildlife. Furthermore, the environmental report failed to consider the potential loss of foraging and feeding grounds for species such as the Iberian Imperial Eagle. It also neglected to mention that the data center will be situated only three kilometers from a critical conservation area for the species. These ecological concerns—particularly those related to non-human impacts—were again overlooked or dismissed in the final assessment. In response, Meta has provoked the creation of the first grassroot organisation against data centres in Spain: *Tu Nube Seca Mi Río* (Your Cloud Dries My River).[46] They considered themselves as a techno-environmental organisation and emerged in April 2023 to:

> “[R]aise awareness about the environmental impact of data centres, especially regarding the use of water resources. We came together as a result of Meta/Facebook’s initiative to create a large data centre in Talavera de la Reina (Castilla y la Mancha, Spain).”
>
> Tu Nube Seca Mi Río.

---

[44] Meta and Zarza Networks, *Proyecto de Singular Interés "Meta Data Center Campus"* (2023), 7-8.

[45] Meta will take 8% of the water allocated to Talavera after reducing its consumption forecast by up to six times: https://toledodiario.es/meta-se-llevara-el-8-de-la-asignada-a-talavera-tras-reducir-hasta-seis-veces-su-prevision-de-consumo/ (Last accessed July 24, 2024).

At an event held in Cambridge (UK), which brought together various local communities across Europe resisting data center developments, a member of a Spanish environmental organization shared her personal journey into activism. She stated that it was Mark Zuckerberg who inadvertently pushed her to become an activist against data centers. Her family had previously resisted the construction of an airport—another so-called singular project backed by the regional government—and she began to investigate the potential negative impacts of Meta’s new data center on the territory. In an area already suffering from severe drought, where farmers are struggling to irrigate their fields, the expansion of water-intensive digital infrastructure raises a pressing question in the context of climate emergency: should we prioritize water for food, or water for the lucrative big tech industry?

### Electricity, pylons, and the sacred hill in Chilean working-class municipality

Huechuraba is a district located on the northern side of Santiago de Chile, notable for its strong working-class heritage. In 1986, residents from the La Pincoya neighborhood in Huechuraba attempted to assassinate Augusto Pinochet, the dictator who ruled Chile from 1973 to 1990.[47] Another defining characteristic of Huechuraba is its close connection and geographical proximity to natural parks. The local community maintains a deep spiritual and traditional bond with its *cerro* (hill), which has been an integral part of their family histories and recreational activities for generations. Historically, parents and grandparents relied on this hill for water and food resources.

---

[46] This grassroot organisation has been actively resisting Meta’s project in Talavera de la Reina: https://tunubesecamirio.com/ (Last accessed July 24, 2024).

[47] Patricia Verdugo and Carmen Hertz, *Operación Siglo XXI* (Santiago de Chile: Catalonia, 2015).

Over the past decade, Huechuraba has faced challenges in providing sufficient housing, leading grassroots organizations to resist and demand more housing from their local council. While residents feel neglected by the wider city of Santiago, their local identity remains strong, rooted in a legacy of resistance and a profound sense of community belonging. Consequently, the people of Huechuraba actively monitor industrial projects proposed within their territory as a political act aimed at protecting their land from industrial developments that fail to serve their interests and threaten their natural environment.

In Chile, the S*ervicio de Evaluación Ambiental* (Environmental Evaluation Service) is a government platform that publicly publishes all environmental reports, offering a transparent and accessible source of environmental information. Through this platform, the Huechuraba community keeps track of proposed projects and their anticipated environmental impacts. It was in this way that they became aware of Amazon's plan to construct a hyperscale data center in their district.

While previous communities in Chile resisted Google's data center primarily due to concerns about water consumption, the case of Amazon in Huechuraba presents a different set of issues. The local community's main concerns focused on the extensive land use required for the infrastructure and the fossil fuel consumption associated with data centers. Additionally, the community discovered through the Environmental Evaluation Service that a separate project—unrelated in official documentation but directly linked to the data center—proposed the construction of 4 kilometers of electrical pylons. This environmental report indicated the installation of 24 pylons across the hills of Huechuraba. Notably, Amazon did not disclose this infrastructure in its own data center environmental report. Instead, the company appeared to employ a "divide and conquer" strategy by fragmenting the environmental impacts into multiple reports: one addressing the data center itself, and another detailing the electrical infrastructure. This fragmentation dilutes the perception of the overall environmental impact. Such a practice, however, may be considered illegal under environmental regulations in Chile.

Similar to the case in Spain, the local community in Huechuraba organized a group called *No a Amazon* ("No to Amazon") to resist the data center project. Over the past two years, this group has taken multiple actions to challenge Amazon's plans. Their first step was to submit formal allegations against Amazon's environmental report, highlighting the fragmentation of the project's environmental impact. To build their case, they carefully analyzed environmental documentation and relevant environmental legislation. Although they do not hold formal legal qualifications, their commitment to defending their territory has led them to dedicate significant personal time—often after work—to prepare these allegations.

The group also invited Amazon representatives to visit Huechuraba. Amazon accepted the invitation, bringing tea and cookies, and presented a list of promised benefits to the community, including claims that the data center would improve services such as Netflix streaming. However, the community critically challenged these assertions, and the meeting ended abruptly when Amazon representatives left the room amid chants of "No To Amazon" from the attendees.

Beyond legal challenges, "No To Amazon" has engaged in extensive community outreach. They designed and distributed informative leaflets across Huechuraba (see Figure 4) and organized public workshops to educate residents and address concerns about the data center. The group also facilitated more creative, community-building activities, such as painting a mural displaying their refusal to this data infrastructural project.

## ***Studying up*** **the Materiality of Artificial Intelligence in the Anthropocene**

In 1972, anthropologist Laura Nader introduced the concept of "studying up", provoking for a methodological and epistemological shift within anthropology to examine how power operates in society posing that: "What if, in reinventing anthropology, anthropologists were to study the colonizers rather than the colonized, the culture of power rather than the culture of the powerless, the culture of affluence rather than the culture of poverty?"[48] Nader criticized that anthropology had historically focused on analyzing the poor, ethnic groups, and disadvantaged communities. However, as she argued, anthropologists should investigate the powerful institutions and their mechanisms affecting society. This framework was later adopted by Diana E. Froysthe who investigated the ethics and politics of science and technology, and more specifically artificial intelligence.[49] Her work consisted in "studying up" engineers developing algorithms by analyzing their values and behaviors. This framework for studying up has also been recently integrated by critical data scholars accounting for "historical inequities, labor conditions, and epistemological standpoints inscribed in data".[50] More concretely, there has been calls to studying up data science which "requires us to develop a critical reflex when presented with opportunities to build models based on data collected by powerful institutions. It requires a new set of reflective practices which push the data scientist to examine the political economy of their research and their own positionality as researchers working in broken social systems."[51] Given the environmental impact of the AI industry, as demonstrated by the cases discussed above, I argue that a "studying up" approach should also be integrated into analyses of AI's materiality.

[48] Laura Nader, "Up the Anthropologist: Perspectives Gained from Studying Up," in *Contrarian Anthropology: The Unwritten Rules of Academia* (New York and Oxford: Berghahn Books, 2018), p.

Within the context of critical data studies in the Anthropocene, *studying up* entails examining structures of power and domination within the political economy of data and artificial intelligence. It requires interrogating who benefits from the extractive logics embedded in the construction and operation of contemporary data infrastructures and to remain politically aware of its supply chains. It also involves questioning why governments are embracing the technological hype promoted by the tech industry and placing trust in the supposed economic benefits of its infrastructures. As academics, rather than fetishizing communities impacted by these systems, we should critically reflect on our own positionality and engage with affected communities to better understand their needs, imaginations, and desires for a world where many worlds fit. This also raises the question: what is our role and responsibility as scholars within these dynamics?

---

[49] Diana E. Forsythe, *Studying Those Who Study Us: An Anthropologist in the World of Artificial Intelligence*, ed. and introd. David J. Hess, Writing Science (Stanford: Stanford University Press, 2001), 119.

[50] Milagros Miceli, Julian Posada, and Tianling Yang, "Studying Up Machine Learning Data: Why Talk About Bias When We Mean Power?" *Proceedings of the ACM on Human-Computer Interaction* 6, no. GROUP (2022): 1.

[51] Chelsea Barabas et al., "Studying Up: Reorienting the Study of Algorithmic Fairness around Issues of Power," in *Proceedings of the 2020 Conference on Fairness, Accountability, and Transparency (FAccT '20)* (Barcelona: ACM, 2020), 170.

The two previous cases illustrate that when *studying up* the materiality of artificial intelligence, the construction of infrastructure often involves practices that obscure its contributions to the Anthropocene. In the Spanish case, environmental organizations have raised concerns that the proposed data center could negatively affect protected fauna. These organizations argue that the environmental impact assessment conducted was superficial and have called for a more thorough and transparent evaluation. Furthermore, the construction of the data center is projected to consume approximately 8% of the municipal water supply—an alarming figure for the local community of Talavera de la Reina, which is already experiencing severe droughts due to climate change.

In the Chilean case, the construction of a data center in Huechuraba has faced resistance from residents, particularly due to the planned installation of high-voltage pylons required to power the facility. Notably, this information became public because Chile was, until recently, the only Latin American country that mandated environmental reports from data centers to be publicly accessible. However, in June 2025, the Chilean government rolled back this legal requirement, effectively deregulating the obligation for data centers to publish their environmental assessments.[52] As of now, both data centers are moving forward with construction, despite documented opposition from the affected communities.

## Conclusion

At the time of writing this chapter during June 2025, southern Europe is experiencing an intense heatwave, with temperatures exceeding 45°C and setting new records in Spain, Italy, and Portugal. Simultaneously, major tech companies have released their annual sustainability reports, acknowledging that the accelerating generative computing race presents significant sustainability challenges due to the vast amounts of energy required to power these technologies. In most cases, this

[52] Paz Peña, *"Brief Comments on the Deregulation of Data Centers in Latin America: The Case of Chile," Instituto Latinoamericano de Terraformación*, June 17, 2025, https://terraforminglatam.net/brief-comments-on-the-deregulation-of-data-centers-in-latin-america-the-case.

energy use translates into carbon emissions, further contributing to Anthropocene-related effects such as global warming. As media scholar Rianne Riemens has investigated, while these reports recognize the environmental impacts of computing and industrial activity, they also reveal contradictions and instances of greenwashing, claims that are increasingly scrutinized in academic literature.[53] Against the backdrop of intensifying climate disruption, this chapter calls for more environmental research by critical data scholars in light of the current climate emergency

This chapter illustrates several mechanisms through which the field of critical data studies could potentially intervene to better understand contemporary datafication within the context of the Anthropocene. First, this chapter provides a theoretical revision of the relationship between datafication, infrastructure, and power, explaining how generative computing is driving the scaling up of resource demand through power mechanisms. Second, the materiality of artificial intelligence highlights how this infrastructure is composed of objects, such as graphics processing units, which require further scrutiny from a critical perspective. This includes identifying the materials and territories exposed to extractivism by the artificial intelligence industry, examining the financial market logics embedded in the politics of infrastructure, and interrogating the power dynamics that underpin it. Third, greenwashing operates as a political strategy that "obfuscates the logics of colonial, capitalist, and power relations that enable [big tech giants'] massive profits".[54] This chapter highlights the diverse environmental impacts associated with data infrastructures. While carbon emissions are often reported in environmental reports, accounting and financial mechanisms frequently obscure their true scale and implications. Other forms of environmental harm, particularly those affecting non-human life, are often excluded from formal assessments. For example, potential threats to protected species such as eagles and vultures in Talavera de la Reina, or to the landscape in Huechuraba, are frequently overlooked. Moreover, as public concern over environmental impacts grows, the Chilean case shows a troubling trend toward increased opacity: the government recently deregulated environmental reporting requirements for data centers, effectively removing public access to critical information. Fourth, this chapter proposes a studying up approach to contemporary datafication and computing infrastructure. This means not only tracing the material and political dimensions of extraction but also investigating the strategies through which the tech industry legitimizes and conceals its environmental impacts while expanding its infrastructure worldwide.

[53] Rianne Riemens, *Platform Earth: Ecomodernism in Tech-on-Climate Discourse* (PhD diss., Radboud University Nijmegen, 2025)

[54] Mél Hogan and Gwendolyn Blue, "Big Cloud Solastalgia," in *Digital Technologies for Sustainable Futures: Promises and Pitfalls*, ed. Chiara Certomà, Fabio Iapaolo, and Federico Martellozzo (London: Routledge, 2025), 185.

It has been ten years since the field of critical data studies was established, calling "for ethnographic and discursive work, for the thick description of data and the cultures around it, just as much as it relies on algorithmic analysis. It is not enough to map Big Data; the point is to change it".[55] Ten years later, I argue that it is still not enough to merely study data. Critical data studies must now engage in political analysis of the Anthropocene, the Capitalocene, the Chthulucene and the Anthrobscene, incorporating debates about the material and infrastructural dimensions embedded in data and algorithms, while advocating for changes that will guide us towards a just and sustainable transformation. This includes "grounding research and analysis in a landscape"[56] to study the materiality of contemporary datafication, while remaining critically conscious about the political influence and power relations of its infrastructure. As critical scholars, we should navigate the logics of extraction by observing forests and rivers as not as resources to be exploited for data and profit, but as vital elements that must be protected. Our ultimate goals should be to serve as a vehicle to guide societies and future generations towards a livable future.

---

[55] Craig M. Dalton, Linnet Taylor, and Jim Thatcher, "Critical Data Studies: A Dialog on Data and Space," *Big Data & Society* 3, no. 1 (2016), 7.

[56] Anna Lowenhaupt Tsing, "When the Things We Study Respond to Each Other: Tools for Unpacking 'the Material,'" in Anthropos and the Material, ed. Penny Harvey, Christian Krohn-Hansen, and Knut G. Nustad (Durham, NC: Duke University Press, 2019), 230.